\documentclass[twocolumns,a4paper]{aa}
\usepackage{graphicx}
\usepackage{txfonts}
\usepackage{color}
\usepackage{natbib}
\usepackage{hyperref}

\usepackage{amsmath}

\definecolor{blue}{RGB}{0,0,255}
\definecolor{red}{RGB}{255,0,0}
\definecolor{green}{RGB}{0,255,0}

\begin{document}

\title{Combining magneto-hydrostatic constraints with Stokes profile inversions. IV. Imposing $\nabla\cdot{\bf B}=0$ condition}
\author{J.M.~Borrero\inst{1} \and A. Pastor Yabar\inst{2} \and B. Ruiz Cobo\inst{3,4}}
\institute{Institut f\"ur Sonnenphysik, Sch\"oneckstr. 6, D-79104, Freiburg, Germany
\and
Institute for Solar Physics, Department of Astronomy, Stockholm University, AlbaNova University 
Centre, 10691 Stockholm, Sweden
\and
Instituto de Astrof{\'\i}sica de Canarias, Avd. V{\'\i}a L\'actea s/n, E-38205, La Laguna, Spain
\and
Departamento de Astrof{\'\i}sica, Universidad de La Laguna, E-38205, La Laguna, Tenerife, Spain
}
\date{Recieved / Accepted }

\abstract{Inferences of the magnetic field in the solar atmosphere by means of spectropolarimetric
inversions (i.e., Stokes inversion codes) yield magnetic fields that are non-solenoidal 
($\nabla\cdot{\bf B} \ne 0$). Because of this, results obtained by such methods are sometimes put into 
question.}{We aim to develop and implement a new technique that can retrieve magnetic fields that are simultaneously
  consistent with observed polarization signals and with the null divergence condition.}{The method used in this work strictly imposes $\nabla\cdot{\bf B}=0$
by determining the vertical component of the magnetic field ($B_{\rm z}$) from the horizontal ones ($B_{\rm x},B_{\rm y}$). 
We implement this solenoidal inversion into the FIRTEZ Stokes inversion code and apply it to spectropolarimetric 
observations of a sunspot observed with the Hinode/SP instrument.}{We show that the solenoidal inversion retrieves
  a vertical component of the magnetic field that is consistent with the vertical component of the magnetic field inferred from the non-solenoidal one.
  We demonstrate that the solenoidal inversion is capable of a better overall fitting to the observed Stokes vector than the non-solenoidal inversion. In fact, the solenoidal magnetic 
field fits Stokes $V$ worse, but this is compensated by a better fit to Stokes $I$. We find a direct correlation between the worsening 
in the fit to the circular polarization profiles by the solenoidal inversion and the deviations in the inferred $B_{\rm z}$ with respect to
the non-solenoidal inversion.}{In spite of being physically preferable, solenoidal 
magnetic fields are topologically very similar in 80\% of the analyzed three-dimensional domain to the non-solenoidal fields obtained 
from spectropolarimetric inversions. These results support the idea that common Stokes inversion techniques fail to reproduce $\nabla\cdot{\bf B}=0$ 
mainly as a consequence of the uncertainties in the determination of the individual components of the magnetic field. In the remaining 20\% of the analyzed
domain, where the $B_{\rm z}$ inferred by the solenoidal and non-solenoidal inversions disagree, it remains to be proven that the solenoidal inversion is 
to be preferred because even though the overall fit to the Stokes parameters improves, the fit to Stokes $V$ worsens. It is in these regions where 
the application of the Stokes inversion constrained by the null divergence condition can yield new insights about the topology of the magnetic field 
in the solar photosphere.}

\titlerunning{MHS constraints in Stokes inversions. IV. $\nabla\cdot{\bf B}=0$}
\authorrunning{Borrero et al.}
\keywords{Sun: sunspots -- Sun: magnetic fields -- Sun: photosphere -- Magnetohydrodynamics
  (MHD) -- Polarization}
\maketitle

\def\kms{~km s$^{-1}$}
\def\deg{^{\circ}}
\def\df{{\rm d}}
\newcommand{\ve}[1]{{\rm\bf {#1}}}
\newcommand{\diff}{{\rm d}}
\newcommand{\Conv}{\mathop{\scalebox{1.5}{\raisebox{-0.2ex}{$\ast$}}}}%
\def\ex{{\bf e_x}}
\def\ez{{\bf e_z}}
\def\ey{{\bf e_y}}
\def\expr{{\bf e_x^\ensuremath{\prime}}}
\def\ezpr{{\bf e_z^\ensuremath{\prime}}}
\def\eypr{{\bf e_y^\ensuremath{\prime}}}
\def\xp{x^\ensuremath{\prime}}
\def\yp{y^\ensuremath{\prime}}
\def\zp{z^\ensuremath{\prime}}
\def\rp{r^\ensuremath{\prime}}
\def\xas{x^{\ast}\!}
\def\yas{y^{\ast}\!}
\def\zas{z^{\ast}\!}
\def\C{\mathcal{C}}

\section{Introduction}
\label{sec:introduction}

Stokes inversion codes are one of the most useful tools to study the magnetic field in the lower solar atmosphere (photosphere and chromosphere). These codes typically
operate by solving the radiative transfer equation for polarized light using an atmospheric model that includes the physical parameters of the plasma (temperature, density, 
magnetic field, etc.) in order to produce a theoretical or synthetic Stokes vector, ${\varmathbb I}^{\rm syn}$, in absorption and emission spectral lines. The synthetic Stokes profiles 
are then compared with the observed one ${\varmathbb I}^{\rm obs}$ and, in case of a mismatch, the physical parameters of the atmospheric model are modified in order to produce 
a better fit to the aforementioned observations. For a review of the different methods available to achieve this, we refer the reader to \citet{jc2016review}.\\

Although originally proposed more than two decades ago by \citet{hector2001review}, in recent years there has been a renewed interest in combining the results from Stokes inversion codes
with physical constraints arising from magnetohydrodynamic (MHD) and Maxwell's equations \citep{puschmann2010pen,tino2017,borrero2019mhs}, as we are now able to retrieve the magnetic field
in the three-dimensional Cartesian domain ${\bf B}(x,y,z )$ in a way that is consistent with the magneto-hydrostatic equilibrium \citep{adur2021,borrero2021mhs}. In spite of
this success, the inferred magnetic field remains inconsistent with Maxwell's second equation that states that the magnetic field must be solenoidal (also known as Gauss' law for the 
magnetic field): $\nabla\cdot{\bf B}=0$. Because of this, the magnetic field inferred from the application of Stokes inversion codes to spectropolarimetric observations is sometimes put into
question. In particular, \citet{horst2018divb} has pointed that Stokes inversion applied to spectropolarimetric data from sunspots yields a vertical variation of $B_{\rm z}$ that is a factor
of five to ten larger than the horizontal variation of $B_{\rm x}$ or $B_{\rm y}$: $dB_{\rm z}/dz \approx -3$ G~km$^{-1}$, whereas $dB_{\rm x}/dx \approx 0.5$ G~km$^{-1}$. On the other hand, some
authors argue that what spectropolarimetry allows to be measured is ${\bf H}$ instead of ${\bf B}$ and that $\nabla\cdot{\bf H}$ is not necessarily zero \citep{bommier2020}. Alternatively,
minimizing the divergence of the magnetic field vector can be used to solve the 180-degree ambiguity in the components of the magnetic field perpendicular to the observer's line of sight
\citep{metcalf1994,metcalf2006} or to establish a common geometrical height scale \citep{loeptien2018zw}.\\

In this work, we put forward the idea that the reason the null divergence condition of the magnetic field, as inferred from spectropolariemtric observations, is not satisfied is 
because of the intrinsic uncertainties in the determination of the different components of the magnetic field. If correct, this interpretation would imply that, within the error margins 
in the inference of $B_{\rm x}$, $B_{\rm y}$, and $B_{\rm z}$, it should be possible to propose a fully solenoidal magnetic field ($\nabla\cdot{\bf B}=0$) that can fit the observed polarization
signals. To test this hypothesis, we applied the FIRTEZ Stokes inversion code \citep{adur2019} to spectropolarimetric observations of a sunspot located at the disk's center observed with
the Hinode/SP instrument (Section~\ref{sec:observations}), and we show that within the errors of the inversion code, we can obtain a solenoidal magnetic field (Section~\ref{sec:bz}) that
is also capable of fitting the observed polarization signals (Section~\ref{sec:fits}) in about 80\% of the analyzed three-dimensional domain. Correlations between the fits to the Stokes
parameters and the inferred magnetic field are studied in Section~\ref{sec:correlations}. The effects of imposing solenoidal magnetic fields in the inference of $\nabla\times{\bf B}$
(i.e., electric currents) is investigated in Section~\ref{sec:currents}. Finally, in Section~\ref{sec:conclusions}, we present our conclusions and implications for the study of magnetic
fields in the solar photosphere via Stokes inversion techniques.\\

\section{Observations and Stokes inversion}
\label{sec:observations}

\begin{figure}
\begin{center}
\includegraphics[width=8cm]{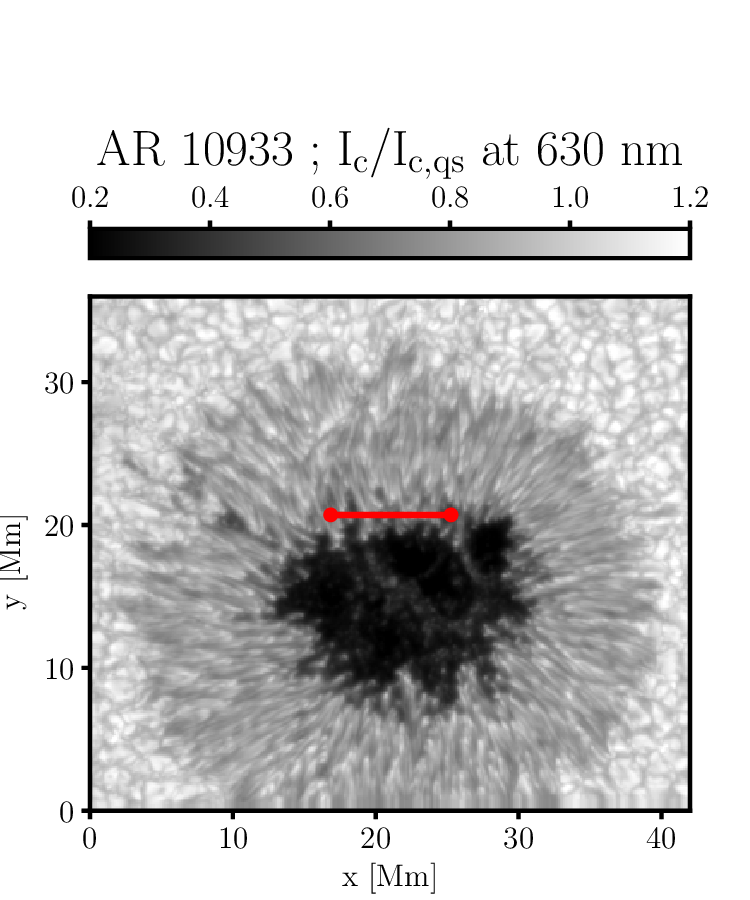}
\end{center}
\caption{Map of the continuum intensity at 630 nm in AR 10933 observed at disk center on January 6, 2007, by SP on board the Hinode spacecraft.
Physical parameters on the $(x,z)$ slice indicated by the horizontal red line are studied in Section~\ref{sec:bz}.\label{fig:icmap}}
\end{figure}

In this work, we analyze the sunspot AR 10933 observed at disk center on January 6, 2007, between 00:00 and 01:25 UT
with the Hinode spectropolarimeter \citep[SP;][]{lites2001hinode,ichimoto2007hinode}. The Hinode/SP is attached to the Solar Optical Telescope 
\citep[SOT;][]{suematsu2008hinode,tsuneta2008hinode,shimizu2008hinode} on board the Japanese satellite 
Hinode \citep{kosugi2007hinode}. The observed data comprise the Stokes vector ${\varmathbb I}^{\rm obs}=(I,Q,U,V)$ in a spectral region around 630 nm that contains two 
Fe {\sc I} lines across a total of $N_\lambda=112$ wavelength positions with a wavelength sampling of about 21.5 m{\AA}. The atomic parameters for these spectral lines 
can be found in \cite{borrero2014milne} (see their Table 1). The SP is a slit-spectrograph where a given region is scanned spatially. For each slit
position, the light is integrated for a total of 4.8 seconds, yielding a photon noise level of about $\sigma = 10^{-3}$ in units of the quiet Sun continuum intensity. A 
map of the continuum intensity $I_{\rm c}$, normalized to the average quiet Sun continuum intensity $I_{\rm c,qs}$, can be seen in Fig.~\ref{fig:icmap}. This map 
includes $n_{\rm x}=350$ and $n_{\rm y}=300$ pixels on the horizontal $(x,y)$ plane, with an spatial sampling of about 0.16 arcsec (i.e., $\df x = \df y = 120$~km at disk center).\\

The observed Stokes vector is analyzed using the FIRTEZ Stokes inversion code \citep{adur2019} with the following number of free parameters: eight for the temperature $T$, eight for
the line-of-sight velocity (i.e., or vertical velocity at disk center) $v_z$, and four for each of the three components of the magnetic field $B_x$, $B_y$, and $B_z$, for
a total of $F=28$ free parameters. The free parameters are located equidistantly along the $z$ direction, which in turn is discretized along $n_{\rm z}=128$ points. The grid 
size along this direction is $\df z=12$~km. During the inversion, all four Stokes parameters are given the same weight ($w_{\rm i}=w_{\rm q}=w_{\rm u}=w_{\rm v}=1$) in the 
definition of the $\chi^2$ merit function that is being minimized:\\

\begin{equation}
\chi^2 = \frac{1}{4 N_\lambda-F} \sum\limits_{k=1}^{4}\sum\limits_{j=1}^{N_\lambda} [\varmathbb{I}^{\rm obs}_{k}(\lambda_j)-\varmathbb{I}^{\rm syn}_{k}(\lambda_j)]^2 \left( \frac{w_{k}}{\sigma} \right)^2  \;,
\label{eq:chi}
\end{equation}

where the denominator $4 N_\lambda-F$ refers to the number of degrees of freedom, that is, the difference between the number of data points and the number of free parameters.
The index $k$ refers to each of the four Stokes parameters $I$, $Q$, $U$, and $V$ so that $\varmathbb{I}_1 = I$, $\varmathbb{I}_2=Q$, and so forth. The term $\sigma$ refers to the photon noise.  
We note that in this work, we are not performing any regularization, and therefore no additional
terms appear in Eq.~\ref{eq:chi}. As a result of the inversion, FIRTEZ provides the aforementioned physical parameters in the 
three-dimensional Cartesian domain $(x,y,z )$. This is possible because FIRTEZ calculates the gas pressure $P_{\rm g}$ employing magneto-hydrostatic equilibrium instead of 
hydrostatic equilibrium \citep{borrero2019mhs,borrero2021mhs}. This is done after the Stokes inversion runs at every $(x,y)$ pixel by the MHS module that iteratively solves the 
following Poisson-like equation:\\ 

\begin{equation}
\nabla^2 (\ln P_{\rm g}) =  -\frac{u g}{K_b} \frac{\partial}{\partial z}\left[\frac{\mu}{T}\right] + 
\frac{1}{c} \nabla\cdot\left[\frac{{\bf j} \times {\bf B}}{P_g}\right] \; .
\label{eq:mhsnew}
\end{equation}

We note that FIRTEZ has an internal implementation of the non-potential field calculation method \citep{manolis2005}, so it provides the magnetic field without the 180$^{\circ}$ ambiguity
in the $B_x$ and $B_y$ components of the magnetic field. The inversion performed under the conditions described above will be referred to as MHS (or simply as non-solenoidal) inversion, and the inferred magnetic field 
will be referred to as ${\bf B}_{\rm MHS}$. A graphical illustration of how the FIRTEZ code operates can be found in Figure 2 in \citet{borrero2021mhs}. Following this inversion, we performed 
a second inversion using as the initial guess the results from the MHS inversion, where $B_{\rm z}$ is no longer inverted, thus having only $F=24$ free parameters. However, the Stokes $V$ was still 
fitted: $w_{\rm v}=1$. Instead of inferring $B_{\rm z}$ from the Stokes inversion, it was obtained by considering that the magnetic field must be solenoidal:

\begin{equation}
\frac{\partial B_{\rm z}}{\partial z} = - \left(\frac{\partial B_{\rm x}}{\partial x}+\frac{\partial B_{\rm y}}{\partial y} \right) \;.
\label{eq:divb}
\end{equation}

We numerically solved this differential equation via finite differences, where the spatial derivatives were approximated with a second-order centered difference formula
for the inner grid points along the $z$ direction and using second-order forward and backward difference formulas at the boundaries. Given $B_{\rm x}$ and $B_{\rm y}$, this led to a linear system
of equations that could be easily solved via a matrix inversion, yielding $B_{\rm z}(z)$. The process was then repeated for every $(x,y)$ point, thereby giving $B_{\rm z}(x,y,z )$.
We note that since Eq.~\ref{eq:divb} is a first-order differential equation, only one boundary condition was needed. We selected neither the upper 
($z=z_{\rm max}=n_z*\df z=1536$~km) nor the lower ($z=0$) boundaries because these are typically located in a region where the spectral lines carry no useful information about 
the magnetic field. Instead, we used as a boundary condition the value of the vertical component of the magnetic field from the previous MHS (non-solenoidal) inversion at a location where
the employed spectral lines observed by Hinode/SP provide the largest sensitivity: $B_{\rm z,MHS}(\log\tau_{\rm c}=-1.5)$. From there, we performed two integrations of Eq.~\ref{eq:divb}: one downward 
from $z(\log\tau_{\rm c}=-1.5)$ until $z=0$ and another upward from $z(\log\tau_{\rm c}=-1.5)$ until $z=z_{\rm max}$. Because at $z(\log\tau_{\rm c}=-1.5)$ we employed second-order
forward or backward finite differences, which are known to be slightly less accurate than centered differences, the accuracy to which $\nabla\cdot{\bf B}=0$ can be achieved
at this height is slightly worse than above and below this point: $\|\nabla\cdot{\bf B}\| \approx 10^{-4}$ vs $\|\nabla\cdot{\bf B}\| \approx 10^{-6}$. Next, we note that according to
Eq.~\ref{eq:divb}, once $B_{\rm z}(z)$ is obtained, any value can be added to $B_{\rm z}(z)$ as long as it is constant with $z$. We took advantage of this fact to further refine
$B_{\rm z}(z)$ so that its average value between $\log\tau_{\rm c}=[0,3]$ (i.e., the region where the line is formed) is the same as in the MHS inversion. Together with our selection of
the boundary condition, this ensured that the Stokes $V$ signals produced by the newly obtained $B_{\rm z}(z)$ did not deviate too much from the observed ones since $B_{\rm z,MHS}(z)$ had
been previously obtained in order to fit the observed Stokes $V$ signals.\\

\begin{figure}
\begin{center}
\includegraphics[width=8cm]{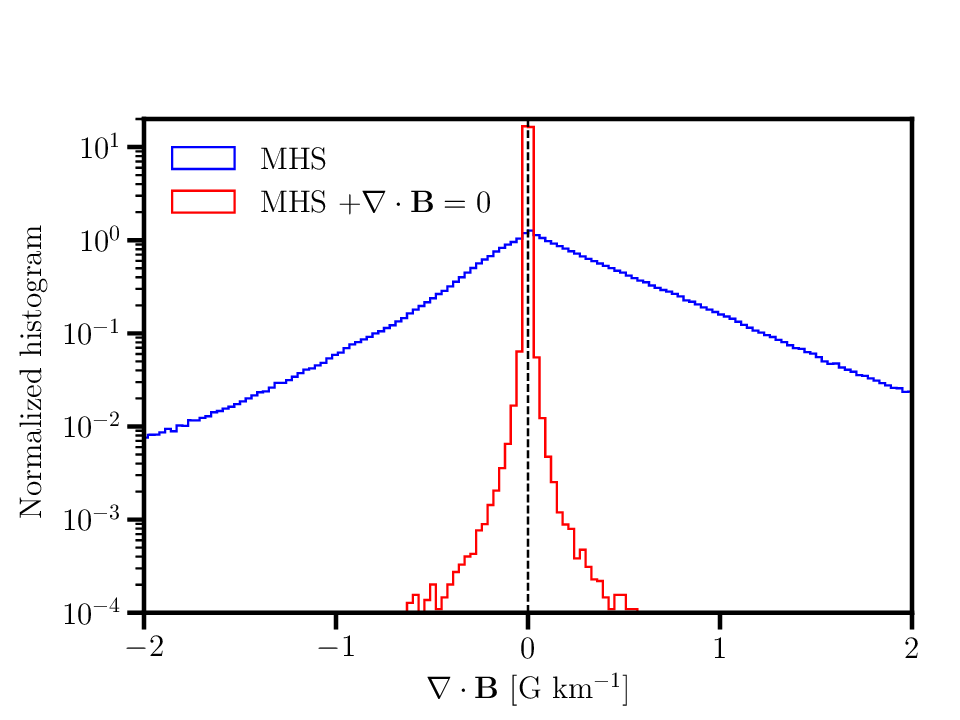}
\end{center}
\caption{Histograms of the divergence of the magnetic field from the first MHS inversion (blue) and second MHS + $\nabla\cdot{\bf B}=0$ inversion (red).\label{fig:divb_histogram}}
\end{figure}

The calculation of $B_{\rm z}$ in the fashion described above does not take place at every iteration step of the Stokes inversion itself but rather at the end, once it has been run for
all $(x,y)$ pixels. This is done because it is only at this stage that the 180-degree ambiguity in $(B_{\rm x},B_{\rm y})$ has been corrected \citep[see Fig.~2 in][]{borrero2021mhs} and that 
$B_{\rm x}(x,y,z )$ and $B_{\rm y}(x,y,z )$ are known, thus allowing the right-hand side of Eq.\ref{eq:divb} to be determined. This choice has the added benefit that the MHS module receives a 
fully solenoidal magnetic field, and therefore, the electric currents (${\bf j} \propto \nabla\times{\bf B}$) and the Lorentz force (${\bf L} \propto {\bf j} \times {\bf B}$ ) used to 
establish the magnetic-hydrostatic equilibrium arise from a more realistic magnetic field (see also Section~\ref{sec:currents}). Now, owing to the existence of several cycles between the 
Stokes inversion module and the MHS module, $B_{\rm z}(z)$ is adjusted several times to reflect the possible changes in $B_{\rm x}$ and $B_{\rm y}$ that might occur and that affect the right-hand 
side of Eq.~\ref{eq:divb}. If the linear polarization signals ($Q$ and $U$) are strong, convergence in $B_{\rm z}(z)$ is achieved after only one or two cycles between the Stokes inversion
and the MHS module. If $Q$ and $U$ are weak, convergence cannot be ensured because $B_{\rm x}$ and $B_{\rm y}$ obtained from the Stokes inversion change 
substantially after each cycle, and therefore the right-hand-side of Eq.~\ref{eq:divb} changes somewhat randomly. In these cases, our choice of boundary condition still ensures that the average value $B_{\rm z}$ over the region
where the spectral line is formed stays the same, but the gradient $\partial B_{\rm z} / \partial z$ is not well constrained. The second inversion described here is hereafter referred to 
as MHS + $\nabla\cdot{\bf B}=0$ or simply as "solenoidal inversion."\\

\begin{figure}
\begin{center}
\includegraphics[width=8cm]{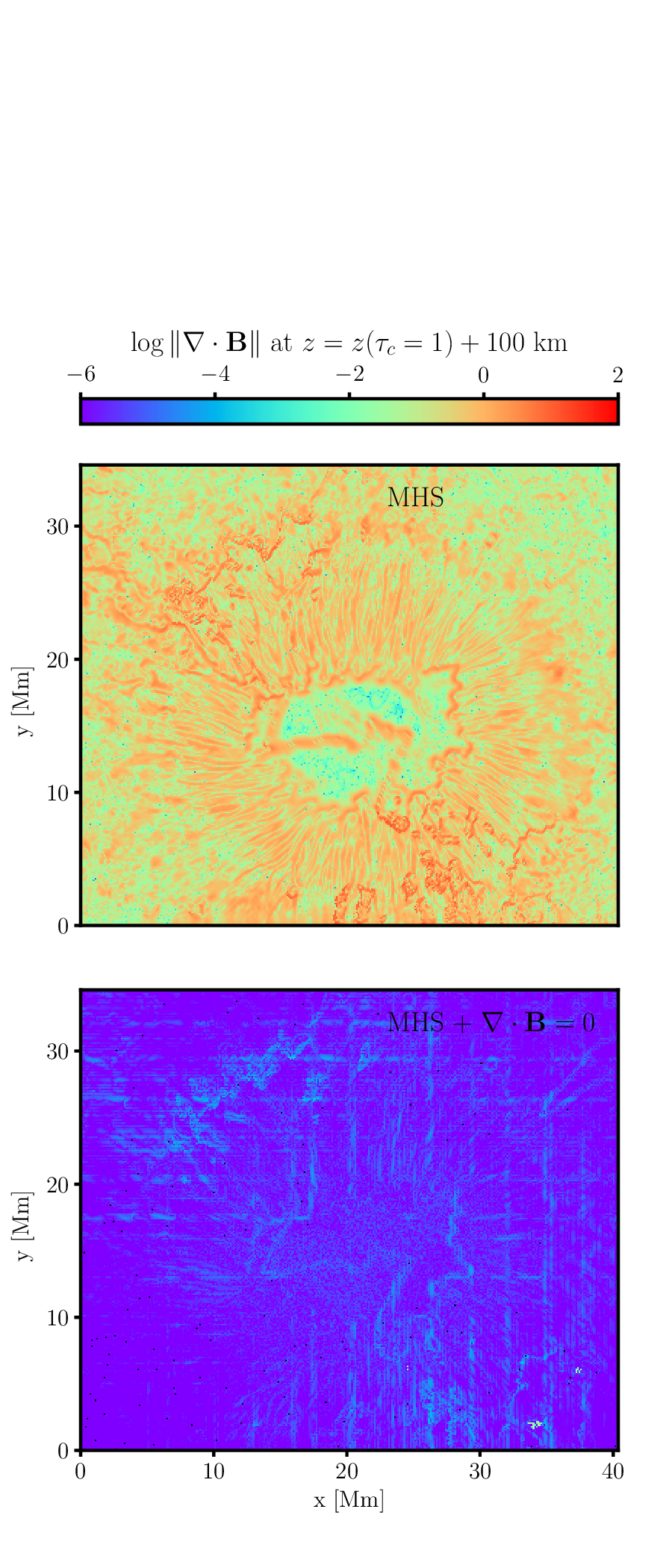}
\end{center}
\caption{Horizontal $(x,y)$ maps of the logarithm of the absolute value of the divergence of the magnetic field at a height of 100 km above the continuum forming layer.
Top panel: Results from the first MHS inversion. Bottom panel: Results from the second MHS + $\nabla\cdot{\bf B}=0$ inversion. Units of $\nabla\cdot{\bf B}$ are in Gauss per kilometer.
\label{fig:divb_maps}}
\end{figure}

As proof that the magnetic field obtained by the second inversion is indeed solenoidal, we present in Figure~\ref{fig:divb_histogram} the histogram of $\nabla\cdot{\bf B}$
for the first MHS inversion (blue) and the second MHS + $\nabla\cdot{\bf B}=0$ inversion (red). As can be seen, the second inversion retrieves a magnetic field that satisfies 
the null divergence condition up to the numerical precision given by the order of the finite difference used to solve Eq.~\ref{eq:divb}. A logarithmic scale is used because the red curve
appears as an almost perfect $\delta$-Dirac in a linear scale. We note that unlike \citet{puschmann2010pen} and \citep{loeptien2018zw}, we do not minimize $\nabla\cdot{\bf B}$ but rather
strictly impose it. Figure~\ref{fig:divb_maps} illustrates the spatial distribution of $\|\nabla\cdot{\bf B}\|$ across the sunspot for the MHS (non-solenoidal) inversion (top) and the
solenoidal inversion (bottom) at a height of 100 km above the height where the continuum is formed: $z=z(\tau_{\rm c}=1)+100$~km. As one can see, the value of the divergence of the magnetic
field vector in the second inversion is about five to six orders of magnitude smaller than in the first inversion. Similar results were obtained at different heights as well.\\

\section{Results: Inference of $B_{\rm z}(z)$}
\label{sec:bz}

\begin{figure}
\begin{center}
\includegraphics[width=8cm]{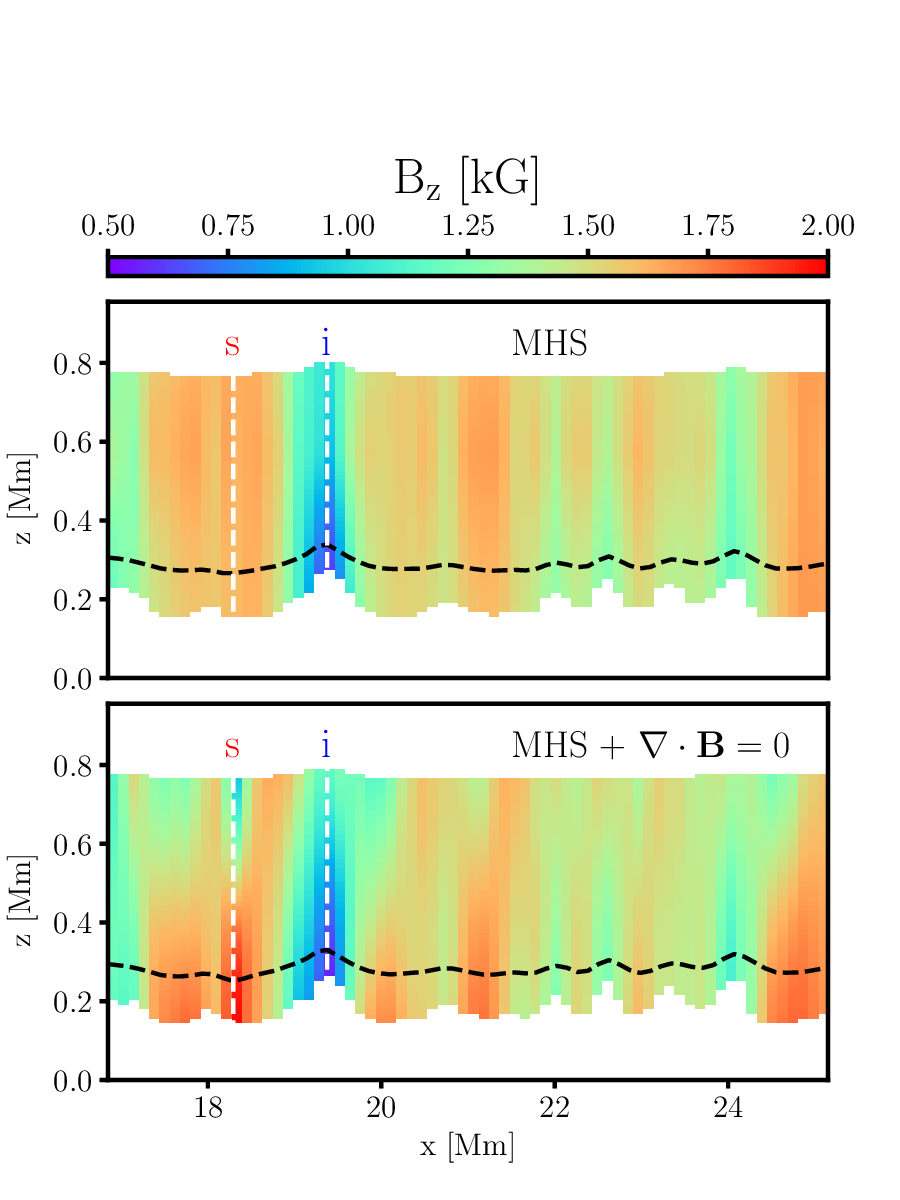}
\end{center}
\caption{Vertical component of the magnetic field, $B_{\rm z}(x,z)$, along the slice indicated by the red line in Fig.~\ref{fig:icmap}. Only regions
between $\log\tau_{\rm c} \in [1,-4]$ (approximate region of sensitivity of the spectral lines observed by Hinode/SP) are shown. The dashed black line
shows the location of the $z(\tau_{\rm c}=1)$ level (i.e., Wilson depression). Results from the MHS inversion are shown in the top panel, whereas
results from the MHS + $\nabla\cdot{\bf B}=0$ inversion are displayed in the bottom panel.\label{fig:vertical_slice}}
\end{figure}

In order to compare the topology of the magnetic field, in particular of $B_{\rm z}$ since this is the one that the solenoidal inversion
determines differently from the MHS (non-solenoidal) inversion, we present in Figure~\ref{fig:vertical_slice} the vertical component of the magnetic field
in the $(x,z)$ plane for the slice indicated by the red line in Fig.~\ref{fig:icmap}. The reason for choosing this region is because it is
in the penumbra, where the magnetic field of a sunspot is most inhomogeneous, and therefore it is here that the two inversions performed in this
work could potentially lead to more different results. Interestingly, both inversions show the well-known spine and intraspine
structure of the penumbral magnetic field \citep{lites1993,valentin1997,borrero2008}, whereby weak horizontal fields (i.e., intraspines) are 
embedded in stronger and more vertical fields (i.e., spines). This is also often referred to as the uncombed penumbral structure 
\citep{solanki1993,borrero2007,tiwari2013}. In this case, the inclusion of the constraints imposed by Gauss' law for magnetism does not lead
to significant changes in the overall structure of the magnetic field.\\

\begin{figure}
\begin{center}
\includegraphics[width=8cm]{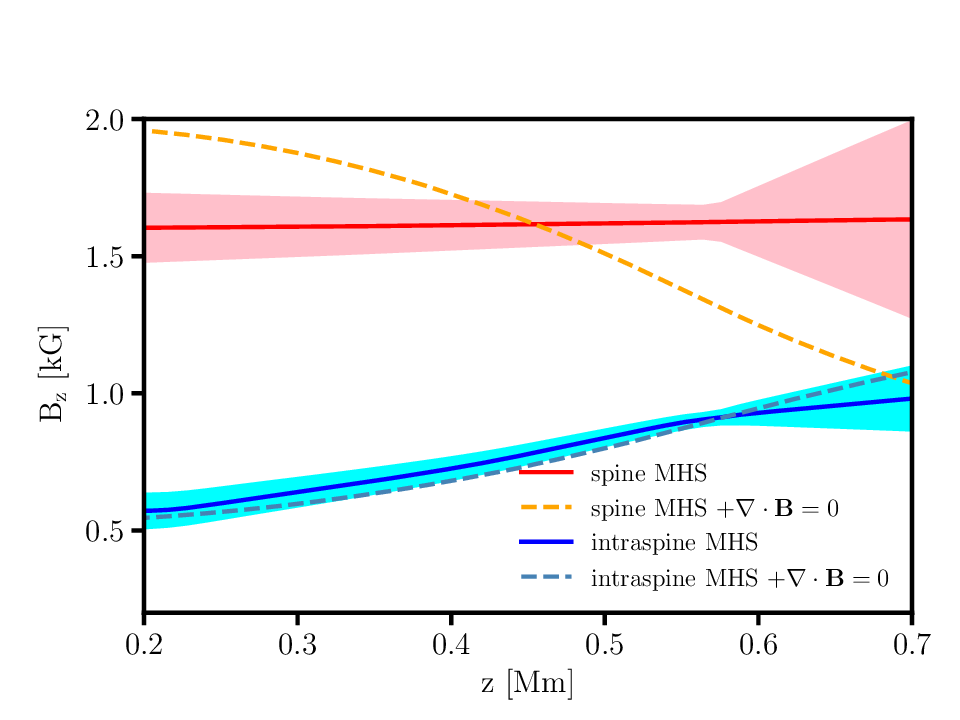}
\end{center}
\caption{Vertical stratification of the vertical component of the magnetic field, $B_{\rm z}(z)$, along the two positions in Fig.~\ref{fig:vertical_slice} (see vertical white
dashed lines). The red and orange colors correspond to the results along the spine from the MHS and MHS + $\nabla\cdot{\bf B}=0$ inversion, respectively. The shaded pink areas indicate
the $3\sigma_{\rm B_{\rm z}}$ confidence level from the MHS inversion. Similarly, the blue and steel blue colors correspond to the results along the intraspine from the MHS and MHS + $\nabla\cdot{\bf B}=0$ 
inversion, respectively. The shaded cyan areas indicate the $3\sigma_{\rm B_{\rm z}}$ confidence level from the MHS inversion.\label{fig:bz}}
\end{figure}

A more detailed comparison can be carried out by focusing on two positions along the $x$ axis that roughly correspond to the locations of the center
of a spine ($x \approx 18.3$~Mm) and the center of an intraspine ($x \approx 19.5$~Mm). In Fig.~\ref{fig:vertical_slice}, these two locations are marked by the vertical
dashed white lines and labeled as \emph{s} (red; spine) and \emph{i} (blue; intraspine). Figure~\ref{fig:bz} displays
$B_{\rm z}(z)$ for these two locations. Along the intraspine, both the MHS inversion (solid blue) and solenoidal inversion (dashed steel blue) yield 
$dB_{\rm z}/dz > 0$. In fact, both results agree extremely well within the 3$\sigma$ error bars from the MHS inversion (shaded cyan area). We emphasize here 
that FIRTEZ finds the standard deviation $\sigma$ in the inference of the different physical parameters by diagonalizing the modified Hessian 
matrix $\varmathbb{H}^{'}$ corresponding to the atmospheric model that best fits the observed Stokes vector. For instance, for a physical parameter denoted by 
index $j$ \citep[see Appendix B or Chapter 11.2.1 in ][ respectively]{jorge1997,jc2003book},

\begin{equation}
\sigma_j^2 = \frac{\chi^2}{F} (\varmathbb{H}^{'-1})_{jj} \;,
\label{eq:sigma}
\end{equation}

where the modified Hessian matrix $\varmathbb{H}^{'}$ is nearly the same as the regular Hessian matrix except that the diagonal elements are 
multiplied by a variable factor that controls whether the inversion is far or close to a minimum, as prescribed by the Levenberg-Marquardt algorithm \citep{levenberg1944,press1986num}.
In the case of the spine, the MHS (non-solenoidal) inversion yields a vertical component of the magnetic field $B_{\rm z}$ that is almost constant with height (solid red line 
in Fig.\ref{fig:bz}), whereas the solenoidal inversion retrieves a $B_{\rm z}(z)$ that decreases by about 1000 Gauss in 500 kilometers (dashed orange line) and lies 
well beyond the 3$\sigma$ error bars from the MHS inversion. We note that the vertical gradient in this case is 
about $dB_{\rm z}/dz \approx -2$ G~km$^{-1}$. This large vertical gradient has been deemed to be at odds with the $\nabla\cdot{\bf B}$ condition for the magnetic field 
\citep{horst2018divb}, and this is probably correct if it should be present through the entire penumbra. However, our inversion demonstrates that locally, this large 
gradient is not only consistent with but actually demanded by the null divergence condition of the magnetic field.\\

\begin{figure}
\begin{center}
\includegraphics[width=8cm]{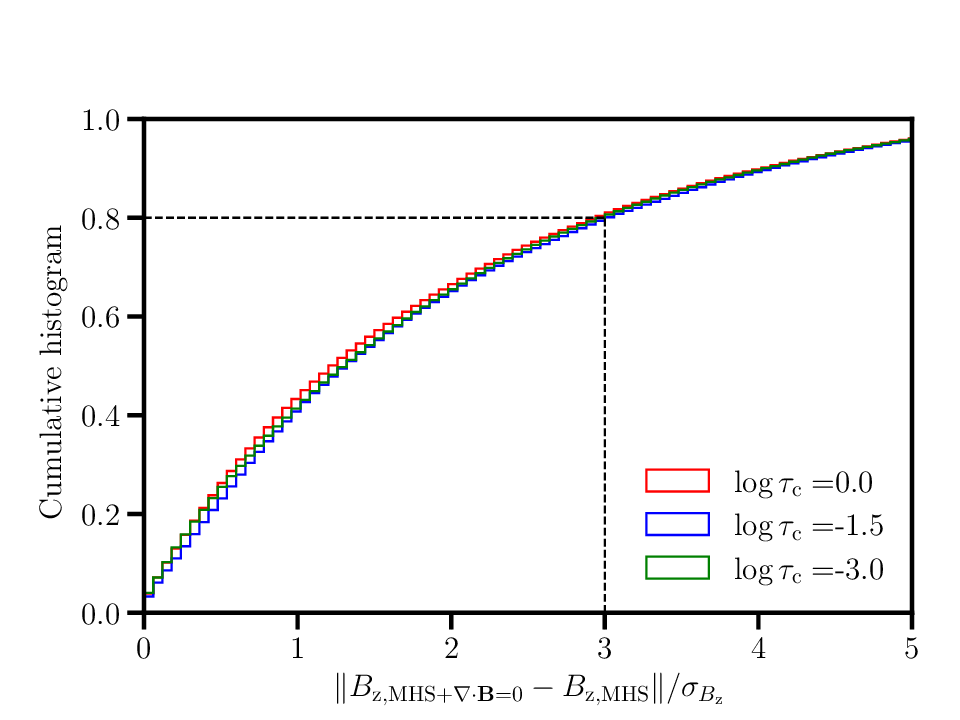}
\end{center}
\caption{Cumulative histogram of the number of pixels, shown at three different optical depth levels (red: $\log\tau_{\rm c}=0$; blue: $\log\tau_{\rm c}=-1.5$; green: $\log\tau_{\rm c}=-3$),
as a function of the ratio of the difference between $B_{\rm z}$ obtained from both inversions to the standard deviation $\sigma_{B_{\rm z}}$. Only pixels where $B_{\rm h}(\log\tau_{\rm c}=-1.5) > 300$~G
are considered (see text for details).\label{fig:sigmabz}}
\end{figure}

Based on the results from the previous two examples (spine and intraspine), we subsequently aimed at determining how often the vertical component of the magnetic 
field provided by the MHS + $\nabla\cdot{\bf B}=0$ inversion is consistent with $B_{\rm z}$ as inferred 
from the regular MHS (non-solenoidal) inversion. In order to investigate this, we show in Figure~\ref{fig:sigmabz} the cumulative histogram of the grid cells
at different $\log\tau_{\rm c}$ levels as a function of $\Delta B_{\rm z}/\sigma_{B_{\rm z}}$, where  $\sigma_{B_{\rm z}}$ is again the standard deviation
in the determination of the vertical component of the magnetic field in the MHS inversion (Eq.~\ref{eq:sigma}). These histograms take into account only
those regions where the horizontal component of the magnetic field $B_{\rm h}=\sqrt{B_{\rm x}^2+B_{\rm y}^2}$ at $\log\tau_{\rm c}=-1.5$ is larger than 300 Gauss. 
This was done so as to avoid regions where $B_{\rm z}$ is not well constrained via the horizontal magnetic fields (see discussion in Sect.~\ref{sec:observations}).
Figure~\ref{fig:sigmabz} shows that in 80\% of the three-dimensional $(x,y,z )$ domain, $B_{\rm z}$ from the solenoidal inversion is within $B_{\rm z} \pm 3\sigma_{B_{\rm z}}$
of the regular MHS (non-solenoidal) inversion. From a statistical point
of view, we can interpret this result in the following way: The magnetic field inferred from the solenoidal inversion is, overall,
similar to the magnetic field provided by the non-solenoidal inversion, and therefore previous results obtained regarding the topology of the
magnetic field in sunspots based on the inversion of spectropolarimetric data \citep[see][for a review of those results]{borrero2011} will likely
stand. However, the $3\sigma$ tolerance is large enough, and the fraction of the three-dimensional domain where this is not satisfied is also large enough ($\approx 20\%$) that 
this statement should be confirmed in a future work and in a case-by-case basis.\\

\section{Results: Fits to the polarimetric signals}
\label{sec:fits}

In order to accept the results from the MHS + $\nabla\cdot{\bf B}=0$ inversion as reliable, it is customary to show that they can fit the polarization
signals to a degree that is at least comparable to the fits produced by the regular MHS inversion. Otherwise, whatever physical realism we might gain
from including the constraint imposed by Maxwell's equation will be defeated by the fact that it is unable to explain the observed Stokes vector.\\ 

\begin{figure*}
\begin{tabular}{cc}
\includegraphics[width=8cm]{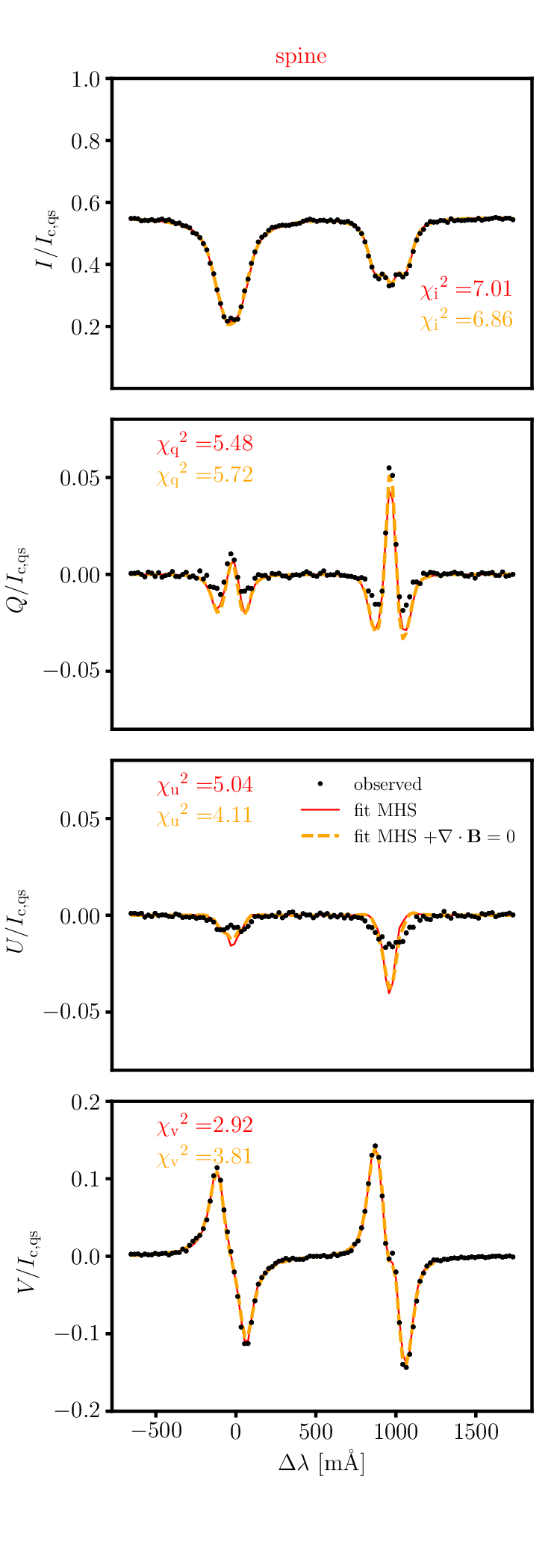} &
\includegraphics[width=8cm]{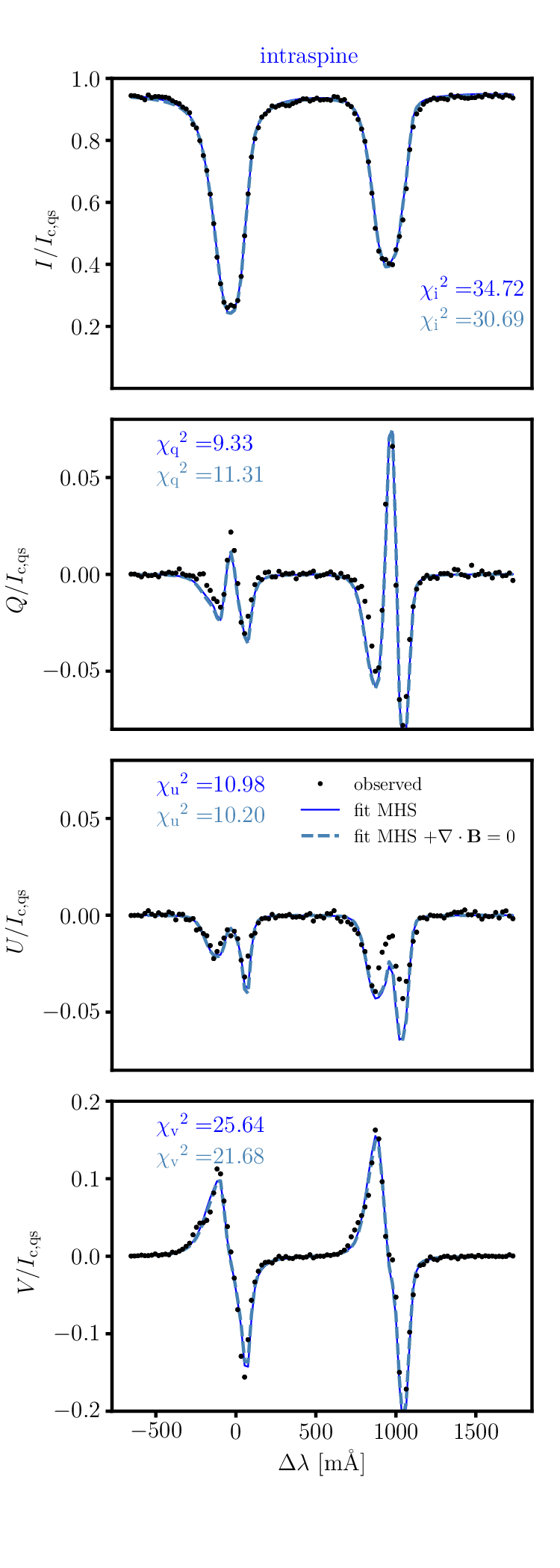}
\end{tabular}
\caption{Comparison between observed and fitted Stokes profiles. Left panels: Observations (black circles) and fits (color lines) at the location of the spine. Right panels:
  Observations (black circles) and fits (color lines) at the location of the intraspine. From top to bottom: Stokes $I$, $Q$, $U$, and $V$. Solid color lines (red and blue) are
  the fits produced by the MHS inversion, whereas dashed color lines (orange and steel blue) correspond to the fits produced by the MHS $\nabla\cdot{\bf B}=0$ inversion. The $\chi^2$ values
for each case are also provided. The wavelength scale is given with respect to the first Fe {\sc I} line at $\lambda_0 = 6301.5012$ {\AA}.\label{fig:fits}}
\end{figure*}

To answer this question, in the context of the selected spine and intraspine, we present in Figure~\ref{fig:fits} the observed (circles) and best-fit profiles (color lines) for both MHS
and solenoidal inversions. In addition, we also provide the $\chi^2$ value for each Stokes parameter separately. At first glance it is clear that both inversions fit the observed Stokes 
profiles very well and that the fits are almost indistinguishable for both kinds of inversions. We emphasize, however, that 
$B_{\rm z}$ was not being explicitly inverted in the case of the MHS +$\nabla\cdot{\bf B}=0$ inversion (see Sect.~\ref{sec:observations}), and therefore, it is somewhat surprising 
that Stokes $V$ is fitted with such accuracy also in this inversion. Perhaps this was to be expected in the case of the intraspine (blue and steel blue lines in the right panels of Fig.~\ref{fig:fits}) 
because $B_{\rm z}(z)$ was almost identical in the MHS and solenoidal inversions (see Fig.~\ref{fig:bz}), but it comes as a surprise that, qualitatively, the fits to Stokes $V$ are also so close
in the case of the spine (red and orange lines in the left panels in Fig.~\ref{fig:fits}) since both inversions feature a rather different $B_{\rm z}(z)$, that is, they differ beyond the 
$3\sigma$ level. This has been possible thanks to our choice of boundary conditions when integrating Eq.~\ref{eq:divb} (see discussion on Sect.~\ref{sec:observations}). We note, however, that
quantitatively the fits to Stokes $V$ are slightly worse in the solenoidal inversion than in the MHS (non-solenoidal) inversion: $\chi_{\rm v} = 3.81$ vs. $\chi_{\rm v} = 2.92$, respectively.\\

\begin{figure}
\includegraphics[width=8cm]{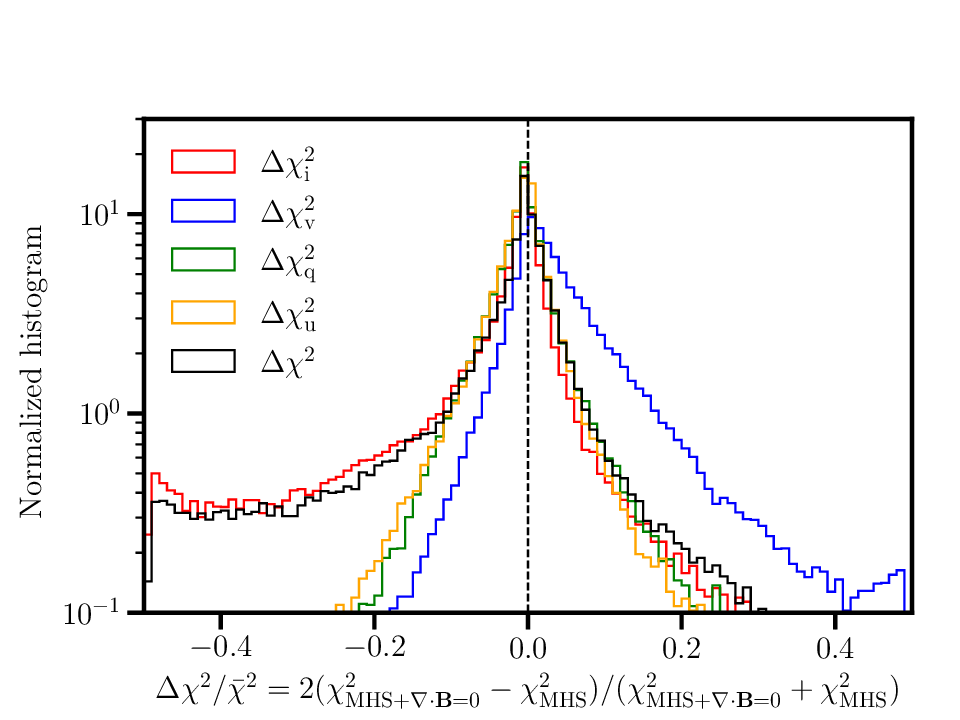}
\caption{Histogram of the differences of the $\chi^2$ normalized to the average $\bar{\chi}^2$ achieved by the MHS+$\nabla\cdot{\bf B}=0$ inversion and by the MHS inversion. 
  Color lines refer to Stokes $I$ (red), Stokes $V$ (blue), Stokes $Q$ (green), Stokes $U$ (orange), and all four Stokes parameters (black). Only pixels where
  $B_{\rm h}(\log\tau_{\rm c}=-1.5) > 300$~G are considered (see text for details).\label{fig:chi}}
\end{figure}

To investigate whether the results for the two aforementioned examples also apply to the full sunspot, we show in Figure~\ref{fig:chi} histograms of the difference between
the $\chi^2$ achieved by the MHS and the solenoidal inversion. Once more, we restricted ourselves to regions where the horizontal component of the magnetic field $B_{\rm h}(\log\tau_{\rm c}=-1.5)$ is 
larger than 300 Gauss. The plot in the figure shows that, overall, the fit to Stokes $I$ (red) improves after applying the solenoidal inversion. This comes at the expense of a worsening of the fit to Stokes 
$V$ (blue). This was to be expected since $B_{\rm z}(z)$ is not inverted at all during the solenoidal inversion. Regarding the linear 
polarization profiles (Stokes $Q$ and $U$), both inversions fit the observations equally well. Overall, the total $\chi^2$ (black) also improves after application of the
MHS$+\nabla\cdot{\bf B}=0$ inversion, with the mean $\chi^2$ over the entire map being about half of the mean in the MHS inversion. We therefore concluded that, in general, the MHS +$\nabla\cdot{\bf B}=0$ inversion fits the 
observed polarization signals at least with the same degree of accuracy as the MHS inversion in spite of featuring a smaller number of free parameters (i.e., $B_{\rm z}$ is not inverted). Because of these reasons, 
along with the fact that the magnetic field is more physically grounded, one could consider the  MHS +$\nabla\cdot{\bf B}=0$ inversion preferable to the MHS (non-solenoidal) inversion. We emphasize, however, that the 
former is not a substitute for the latter because the MHS inversion is needed in order to initialize the solenoidal one. Had we not used the boundary condition at $B_{\rm z}(\log\tau_{\rm c}=-1.5)$ from the 
MHS inversion and later shifted $B_{\rm z}(z)$ by a constant value to match the average from the MHS inversion between $\log\tau_{\rm c} = [0,-3]$ (see Sect.~\ref{sec:observations}), the solenoidal inversion 
would not have been able, by itself, to fit the observed Stokes profiles.\\

\section{Correlations in $B_{\rm z}(z)$ and $\chi_{\rm v}^2$}
\label{sec:correlations}

In Sect.~\ref{sec:fits}, we showed that the solenoidal inversion yields a fit to Stokes $V$ that is on average worse than the fit by the MHS inversion (see blue line in Fig.~\ref{fig:chi}). As we 
have explained, this was to be expected because in the former inversion, $B_{\rm z}$ is not inverted. In addition to this, in Sect.~\ref{sec:bz}, we showed that in about 20\% of the 
three-dimensional $(x,y,z )$ domain studied, the solenoidal inversion retrieves a vertical component of the magnetic field that is beyond the $3\sigma_{\rm B_{\rm z}}$ uncertainty bars from the MHS inversion. This begs the
question as to whether these two effects are correlated, that is, whether in those regions where $B_{\rm z}(z)$ from the MHS$+\nabla\cdot{\bf B}=0$ inversion differs significantly 
from the $B_{\rm z}(z)$ in the MHS inversion, the fit to the circular polarization by the solenoidal inversion worsens.\\

\begin{figure*}[h!]
\begin{tabular}{cc}
\includegraphics[width=8cm]{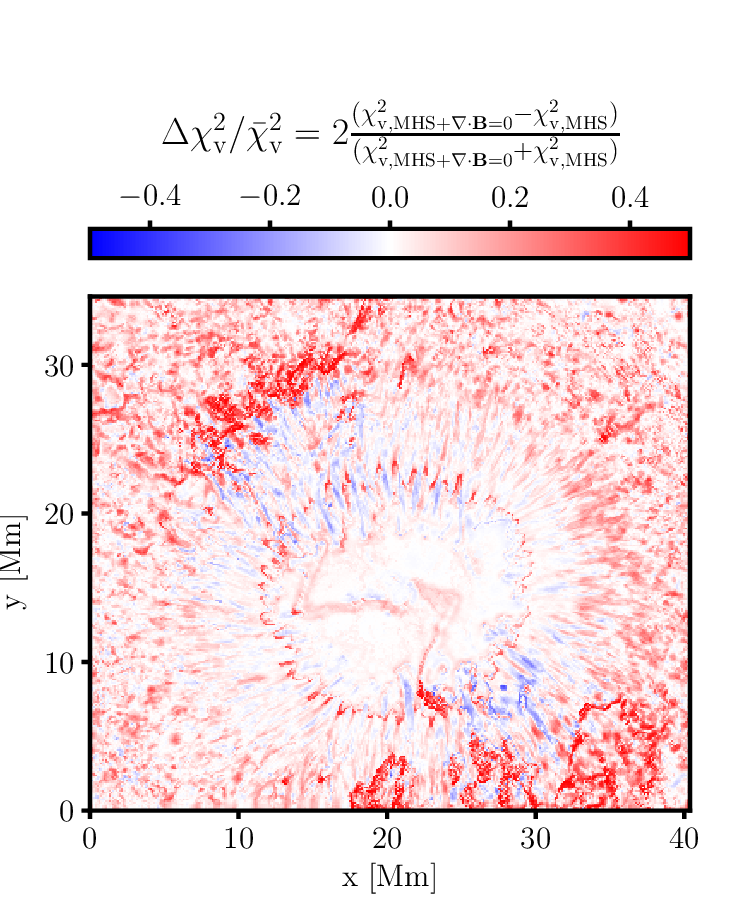} &
\includegraphics[width=8cm]{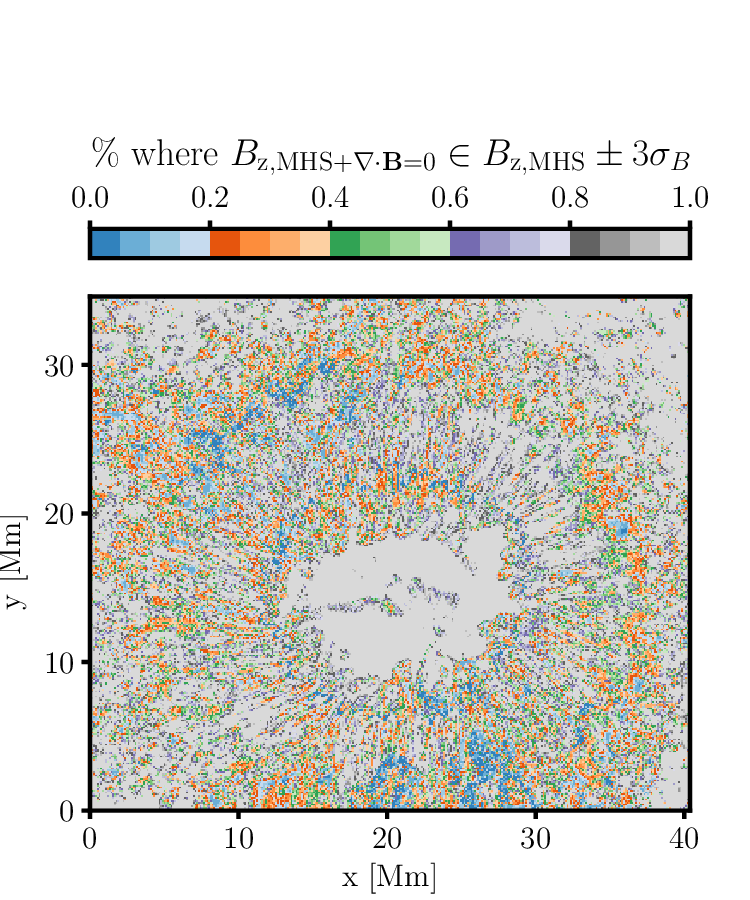}
\end{tabular}
\caption{Correlations between $B_{\rm z}(z)$ and $\chi_{\rm v}^2$. Left panel: Difference in the goodness of the fit to Stokes $V$ (circular polarization) by the solenoidal inversion
  and the MHS inversion normalized to the average of the two. Right panel: Percentage of the region (along the $z$ direction ) where the observed spectral lines are formed in a
  manner in which the vertical component of the magnetic field inferred from the solenoidal and MHS inversions agree with the $3\sigma$ threshold.\label{fig:correlations}}
\end{figure*}

In order to address this question, we present in the left panel of Figure~\ref{fig:correlations}, the spatial distribution of the difference between the $\chi^2$ in Stokes $V$ from
the solenoidal and MHS (non-solenoidal) inversions normalized to the average of the two. Positive regions (red) are those where the MHS inversion fits the circular polarization better. As expected from Fig.~\ref{fig:chi}, positive values should dominate over negative ones. In the right panel of Fig.~\ref{fig:correlations}, we show a map of the sunspot where colors indicate the percentage of pixels along the $z$ direction, but inside the line-formation region ($\log\tau_{\rm c} \in [0,-3]$), where the vertical magnetic field inferred from the MHS$+\nabla\cdot{\bf B}=0$ inversion
falls within $[B_{\rm z,MHS}-3\sigma_{\rm B_{\rm z}},B_{\rm z,MHS}+3\sigma_{\rm B_{\rm z}}]$ in the MHS inversion. The gray regions are those where 80 to 100\% of the grid cells along the vertical 
direction meet this criteria, meaning that $B_{\rm z}(z)$ is very similar in both the solenoidal and MHS (non-solenoidal) inversions. The blue and orange regions are those where the opposite happens.
Comparing the left and right panels in Fig.~\ref{fig:correlations} answers the above question in the affirmative: Regions where $B_{\rm z}(z)$ from the solenoidal inversion differ most from that
of the MHS inversion are the same regions where the fit to the observed Stokes $V$ signals worsens the most. These regions are predominantly located in the outer penumbra, close to the boundary
with the quiet Sun. It is important to emphasize that in spite of producing worse fits to Stokes $V$, the overall $\chi^2$ improves in the MHS$+\nabla\cdot{\bf B}=0$ inversion (see black lines 
in Fig.~\ref{fig:chi}).\\

\section{Electric currents}
\label{sec:currents}

\begin{figure*}[h!]
\begin{center}
\includegraphics[width=17cm]{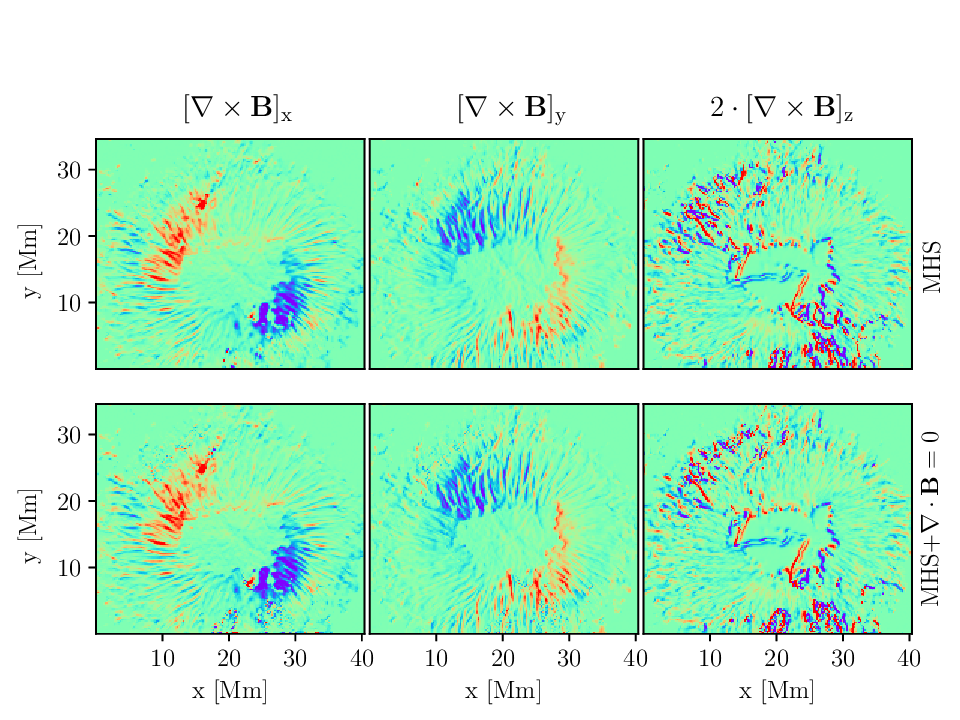}
\end{center}
\caption{Three components of $\nabla\times{\bf B}$ obtained from the MHS inversion (top panels) and MHS + $\nabla\cdot{\bf B}=0$ inversion (bottom panels) at a height that corresponds
to 100 km above the continuum level: $z=z(\tau_{\rm c}=1)+100$~km. For visualization purposes, $[\nabla\times{\bf B}]_{\rm z}$ (right panels) has been multiplied by two. The color scale saturates
at $\pm 4$ G~km$^{-1}$. Only pixels where $B_{\rm h}(\log\tau_{\rm c}=-1.5) > 300$~G are considered (see text for details).\label{fig:currents}}
\end{figure*}

An important question to be addressed here is whether the electric currents as inferred from the regular MHS inversion and
from the MHS + $\nabla\cdot{\bf B}=0$ inversion, and given by $\nabla\times{\bf B}$, differ significantly. In order to do so, we present in Figure~\ref{fig:currents} a comparison of the three components
of the rotational of the magnetic field at $z=z(\tau_{\rm c}=1)+100$~km (i.e., 100 km above the continuum level). These figures show that both inversions yield electric 
currents that are spatially very consistent with each other. Similar levels of consistency were also obtained at other heights, provided that they are located within the
region where the line is formed, namely, $z=z(\tau_{\rm c}=1)$ or $z=z(\tau_{\rm c}=1)+200$~km (not shown). In spite of this spatial consistency, currents determined from each
inversion can present large differences of up to one order of magnitude when looking at individual locations. One might argue that
the electric currents arising from the solenoidal inversion are to be preferred over the electric currents provided by the MHS inversion since they are obtained from
a more physical magnetic field. However it remains to be seen which of the two inversions actually infers the electric currents more reliably. In order to conduct such an investigation, one needs to 
apply the aforementioned inversions to synthetic Stokes profiles produced by MHD simulations where the electric currents are known beforehand \citep{adur2021,borrero2023}.
This, however, falls outside the scope of the present paper.\\

\section{Conclusions }
\label{sec:conclusions}

We have developed a new method that allows Stokes inversion codes for the radiative transfer equation to infer a magnetic field on the solar atmosphere that verifies the null
divergence condition imposed by Maxwell's equation: $\nabla\cdot{\bf B}=0$. Unlike previous methods \citep{puschmann2010pen,loeptien2018zw}, we do not minimize the 
divergence of the magnetic field vector, but instead we strictly impose it. This is done by determining $B_{\rm z}$ from $B_{\rm x}$ and $B_{\rm y}$ such that the magnetic field
is divergence free. The accuracy of the solenoidal solution is only limited by the order of the finite differences formula employed. When applied to spectropolarimetric observations of a 
sunspot recorded with the SP instrument on board Hinode, we easily achieved $\|\nabla\cdot{\bf B}\| \le 10^{-5}-10^{-6}$ G~km$^{-1}$. This method is now implemented and available to users 
of the FIRTEZ Stokes inversion code.\footnote{The FIRTEZ inversion code is open source software, and it is freely available (under GPL 2.0 license) here: \url{https://gitlab.leibniz-kis.de/borrero/firtez-dz-mhs.git}. FIRTEZ was partially developed under the auspices of the Deutsche Forschung Gemeinschaft (DFG project number 321818926)}\\

We have also demonstrated that this new inversion method is capable of fitting the observed Stokes profiles with a very similar degree of fidelity as the inversion method where the magnetic field
is not solenoidal. Interestingly, the vertical component of the magnetic field in the solenoidal case agrees in about 80\% of the analyzed $(x,y,z )$ domain with the non-solenoidal one within the 
$3\sigma$ confidence level regarding the determination of the $B_{\rm z}$. This means that, the overall magnetic topology in the solenoidal inversion is similar to the non-solenoidal one. We interpret these
results as proof that regular (i.e., non-solenoidal) Stokes inversions fail to fulfill the $\nabla\cdot{\bf B}=0$ criterion mostly as a consequence of the uncertainties in the determination
of the individual components of the magnetic field. In about 20\% of the $(x,y,z )$ domain, the solenoidal magnetic field differs from the non-solenoidal one beyond the $3\sigma$ threshold. These deviations
are correlated with a worsening of the ability of the solenoidal inversion to fit the observed Stokes $V$ signals compared to the non-solenoidal inversion. It is in these regions
where our newly developed method can uncover new insights about the topology of the magnetic field. We leave these investigations for a future work.\\

An important limitation of the method described in this paper is that it requires strong polarization signals, mainly in the linear polarization, because $B_{\rm z}$ is obtained
from $B_{\rm x}$ and $B_{\rm y}$. This is certainly the case of sunspots, but it remains to be seen how this method performs in other regions, such as network, internetwork, and plage regions.
Possible improvements to the current implementation of our method might be obtaining a different component of the magnetic field out of the other two or even adjusting which
component is being fixed by the $\nabla\cdot{\bf B}=0$ condition on a pixel-by-pixel basis, depending on the polarization signals observed there.\\

\begin{acknowledgements}
Authors would like to thanks Dr. Ivan Mili\'{c} for suggestions and carefully reading the manuscript. JMB would like to thank K.D.~Leka, H.~Balthasar and V.~Bommier for bringing up to our 
attention, during the SPW9 workshop held in G\"ottingen (Germany) during 2019, some of the issues addressed in this work. BRC has been partially funded by AEI/MCIN/10.13039/501100011033/PID2021-125325OB-C55, 
co-funded by European Regional Development Fund (ERDF) "A way of making Europe". APY is funded/co-funded by the European Union (ERC, MAGHEAT, 101088184). Views and 
opinions expressed are however those of the author(s) only and do not necessarily reflect those of the European Union or the European Research Council. Neither the European Union nor the 
granting authority can be held responsible for them. This research has made use of NASA's Astrophysics Data System. Hinode is a Japanese 
mission developed and launched by ISAS/JAXA, collaborating with NAOJ as a domestic partner, NASA and STFC (UK) as international partners. 
Scientific operation of the Hinode mission is conducted by the Hinode science team organized at ISAS/JAXA. This team mainly consists of scientists from institutes in 
the partner countries. Support for the post-launch operation is provided by JAXA and NAOJ (Japan), STFC (U.K.), NASA, ESA, and NSC (Norway). 
\end{acknowledgements}

\bibliographystyle{aa}
\bibliography{ms}

\end{document}